# Functional analysis of a gene locus in response to non-canonical combinations of transcription factors

Netta Haroush[1], Michal Levo[1], Eric Wieschaus[1,2], and Thomas Gregor[1,3,4]

[1]Lewis-Sigler Institute for Integrative Genomics,
[2]Department of Molecular Biology, Howard Hughes Medical Institute,
[3]Joseph Henry Laboratories of Physics, Princeton University, Princeton, NJ 08544, USA
[4]Department of Stem Cell and Developmental Biology, CNRS UMR3738, Institut Pasteur, 25 rue du Docteur Roux, 75015 Paris, France

**Highlights**

- Highly reproducible patterns emerge despite broad regulator concentration ranges
- In mutants, gene network interactions adjust concentrations towards wild-type levels
- Weak perturbations in regulator concentrations cause precise output pattern shifts
- Strong perturbations generate non-canonical activating input combinations

**Summary**

-Transcription factor combinations determine gene locus activity and thereby cell identity. However, the precise link between concentrations of such activating transcription factors and target-gene activity is ambiguous. Here we investigate this link for the gap gene dependent activation of the *even-skipped* (*eve*) locus in the *Drosophila* embryo. We simultaneously measure the spatiotemporal gap gene concentrations in hemizygous and homozygous gap mutants, and link these to *eve* activity. Although changes in expression extend well beyond the genetically manipulated gene, nearly all expression alternations approximate the canonical combinations of activating levels in wild-type, sometimes necessitating pattern shifts. Expression levels that diverge from the wild-type repertoire still drive locus activation. Specific stripes in the homozygous mutants show partial penetrance, justifying their renown variable phenotypes. However, *all* eve stripes appear at highly reproducible positions, even though a broader span of gap gene expression levels activates *eve*. Our results suggest a correction capacity of the gap gene network and set constraints on the activity of multi-enhancer gene loci.

**Introduction:**

The prevailing model for transcriptional activity of a gene locus postulates that a locus is active when the right combination of transcription factor concentrations is present in the nucleus [1–7]. As individual loci often harbor multiple enhancers, and each enhancer can potentially respond to multiple concentration combinations, distinctly different combinations of transcription factor concentration can activate the same gene [1]. However, it is unknown what the rules and restrictions are for these sets of combinations. How wide is the range of activity-promoting concentrations? How are they related to the distinct activity states of the regulated genes in adjacent cells? How responsive is a locus to novel combinations of regulator concentrations? To shed light on these questions we have manipulated the gap genes in the *Drosophila* embryo to subject the locus of the pair-rule gene *even-skipped* (*eve*) to a range of non-canonical combinations of regulator transcription factor concentrations.

The segmentation patterning program of the early *Drosophila* embryo provides an ideal testbed for quantitative questions pertaining to transcriptional regulation in the functional context of cellular identity formation [8]. A feed-forward flow of information about cellular position starts with primary maternal morphogens that activate the interconnected gap gene network, which in turn drives expression of the pair-rule genes [9]. Pair-rule expression occurs in stripes that are precisely and reproducibly positioned within the embryo, forming an outline for the segmented body plan of the fully developed organism [10]. For this program to result in a precise and reproducible outcome, individual cells need to be instructed and take actions that are consistent with their spatial coordinates within the embryo. At each location along the anterior-posterior (AP) axis, gap gene expression levels contain enough information to determine position with a precision of ~1% egg length (EL), corresponding to the spatial extent of a single cell [11,12]. The resulting pair-rule gene expression can be predicted from the four major gap genes with that same spatial precision [4]. Hence at each position and at the level of an individual cell, the expression levels of the four major gap genes define a four-letter *code word* that instructs gene locus activity.

Here we present a comprehensive data set that establishes the canonical four-letter code word repertoire for the wild-type *eve* locus and quantify the dynamic range over which locus activity ensues. We then challenge the locus' code word repertoire by perturbing the system in hemizygous and homozygous gap gene mutant backgrounds. While the perturbed backgrounds have strong effects on the mean expression profiles of both gap genes and *eve*, the variance in gap gene concentrations and their positional error are very similar to wild-type, as well as the precision of target gene positioning. We show that network induced changes in the concentrations of the three unmanipulated gap genes generally adjust the code word levels to those of the wild-type repertoire, but at displaced locations from the corresponding wild-type Eve stripe positions. Nevertheless, these positions are reproducible and precise, even in homozygous mutants where the penetrance of certain Eve stripes is partial. A largely augmented range of deviations in the homozygous mutants generates code words that can activate multiple distinct stripe enhancers, as suggested by classic studies of reporter construct expression in mutant embryos [13–16]. Using principal component analysis, we identify the major axes of concentration determinants for individual stripes and use the associated novel code words to explain the altered patterns observed in mutant phenotypes.

**Results**

**Precise patterning in hemizygous and homozygous gap gene mutants**

To functionally challenge *eve* locus responses, we alter the code word catalogue beyond its wild-type repertoire. The Drosophila model allows us to accomplish this genetically via null mutations in each of the four major gap genes (i.e., *hunchback* (*hb*), *giant* (*gt*), *Kruppel* (*Kr*), and *knirps* (*kni*)), whose AP patterns overlap nicely with the seven eve stripes which we use as spatial cues for the locus read-out (Figure 1A). Individually for each gap gene, we perform a unique mutation experiment that generates three classes of embryos, all three stemming from the same balancer fly cross. 25% of F1 embryos contain both gene copies (2X, the ones on the balancer chromosome, which we consider as a wild-type equivalent), half of all F1 embryos contain a single copy (1X, hemizygous mutant), and another 25% of F1 embryos contain no copy (0X, homozygous or null mutant) of the wild-type gap gene (Figure 1B). For each gap gene mutation experiment we collect the F1 embryos consisting of these three genotypes and perform immuno-fluorescence labeling for the perturbed gap gene product simultaneously with Eve (our readout). Each embryo was assigned a time stamp with an accuracy of ~1 minute from entry into nuclear cycle 14 (nc14) based on the extent of the progressing cellularization front [17]. To optimally capture the information flow between the gap genes and eve, we focus on *eve* expression around 50 minutes into nc14, allowing an ~8 minutes delay (for changes in gap genes concentration to affect eve levels [18]) from the time point where gap genes produce their maximal information [11]. 2X, 1X, and 0X embryos are sorted based on a classification scheme that acts on gene expression profiles and the associated time stamp (see methods for details).

After measuring Eve, we extract the locations of its maximal activity (peaks) and asses the mutant average displacement from wild-type locations, Δx, and the embryo-to-embryo variability of stripe location, $\delta x$ (Figure 1C). We analyzed over 300 embryos (between 50–100 per experiment), fixed at 48±5 min into nc14.

For each gap gene fly stock, the data set for the 2X balancer homozygotes, faithfully reproduces the wild-type Eve stripe locations and their well-documented spatial precision of ~1% EL (Figure 1D and Figure S1A) [4,10]. Striped expression patterns of Eve are observed in all 1X and 0X mutant embryos (Figure 1E–F and S1B–C), including regions along the AP axis where endogenous expression of the gap gene has been genetically manipulated. For all 1X mutants we observe seven eve stripes, with identities easily defined based on their order along the AP axis, since final differentiation patterns in their hatching larvae are ordered according to wild-type [19–21]. While stripe locations in these embryos are often mildly displaced ($\Delta x \leq 3$ cells) relative to their mean wild-type position (Figure 1E, 1G and Figure S1B), these displaced locations are nevertheless highly reproducible across embryos, in each mutant, with the positional error of any displaced stripe ~1% EL around their shifted mean (Figure 1E, 1I–J, and S1B).

Surprisingly, this precision is also maintained for all Eve stripes in 0X homozygous mutant backgrounds, even though the number of stripes is reduced and variable (Figure 1F, 1H–1J, and S1C). Although stripe identity for these mutants can be ambiguous, many stripes arise in positions that correspond to specific stripes in the wild-type pattern, or very slightly displaced from these positions (>±3 cells, Figure 1F, 1G, and S1C). In regions where the manipulated gap gene would have been expressed in the wild-type, *eve* expression patterns are broadened and consist of a variable number of discrete peaks (Figure S1C–F). Such variable stripes occur with 50–89% frequencies in homozygotes (Figure 1H). When they do occur, however, regardless of their known or unknown identity, their positional error, and that for all stripes in homozygous null mutants, remains at the order of ~1% EL (Figure 1I–J). This result is surprising given the well-documented variable body plan for individual null mutant embryos [20,22]. The gap between the highly reproducible Eve positions and the obvious variability of individual homozygous mutants under identical genetic background can be accounted for by the partial penetrance observed for specific Eve stripes in these mutants. Our data suggest that rather than being a system that has properly evolved under strong evolutionary constraints, spatial patterning precision might be an intrinsic property of the system that reemerges even under novel genetic conditions.

**The underlying basis of eve's pattern reproducibility: quantitative measurement of gap expression in gap mutants**

To investigate the origin of the observed Eve positional precision and associated pattern shifts in the gap gene mutant backgrounds, we examined the twelve different genotypes analyzed for *eve* expression using a protocol that allows simultaneous measurement of protein expression levels for the four gap genes [4,17]. We collected a data set with over 3000 embryos to reconstruct the simultaneous expression dynamics of all four major gap genes for each genotype (Figure 2A–2D, Supplemental Videos 1–4). Due to the interactive nature of the gap gene network, perturbations in gap gene expression levels are not limited to the gap gene whose dosage was genetically manipulated. To quantify primary and secondary effects on gap gene expression, for each genotype we compute the embryo-averaged expression level deviations in space and time of each gap gene from their wild-type counterpart, $\Delta I^g_{gt}(x,t)$ (Figure 2E). We represent these expression level differences in a kymograph where the spatial axis corresponds to the central 80% (or ~54 cells) of the embryo's AP-axis, and the temporal axis corresponds to 10–54±4 min into nc14. We thus construct a total of 48 different kymographs (Figure 2F–I), corresponding to four gap genes (*g*={*hb*, *gt*, *Kr*, *kni*}) measured in 12 genotypes (*gt*={*2X, 1X, and 0X for each for the four gap genes*}). Note that kymographs of the 2x data sets (left columns) serve as a control for the experimental noise in our data measurements (Figure S2A–C & see methods).

As expected, in a given genotype the largest overall deviations, averaged over all time-points and positions (see methods), occur in the gene whose dosage was modified genetically ($\Delta g^g_{1x@g}(x,t)$ and $\Delta g^g_{0x@g}(x,t)$). This overall deviation is ranging between 8–21% and 23–45% of the wild-type maximum, respectively. For all other kymographs, positive (red) and negative (blue) average expression level deviations are comparably weak (ranging between 4-7% for the 1X and 3–16% for the 0X data sets, respectively). The overall absolute expression adjustments of the remaining gap genes in any mutant genotypes are mild compared to the directly perturbed genes. However, these adjustments are higher compared to the randomized control (1–4%), for at least 2 of the remaining genes in homozygous mutants and 1 of the remaining genes in hemizygous mutants. Moreover, although for the unperturbed genes, the averaged deviations over all times and positions is fairly low, their local spatiotemporal deviations (represented by individual pixels in the kymographs) can reach ~25% and ~80% of wild-type peak expression per gene for hemizygous and homozygous mutants, respectively. We observe boundary shifts, identifiable as adjacent red and blue regions in the kymographs (e.g. Gt in *0X-Kr*, or Kr in *1X-hb*), or as pattern expansion (just red, e.g. Kni in *0X-hb*) or pattern contraction (just blue, e.g. Gt in *0X-hb*); as well as expression level increases or decreases without boundary shifts (e.g. Hb in *0X-gt* or Kni in *0X-Kr)*, respectively. Changes in hemizygous mutants either lead to faint boundary shifts or faint expression level decreases.

In what follows, we use the broad range of gap gene combinations as a resource to investigate the downstream impact of gap gene concentrations on *eve* expression. For all our analyses, we choose the time window around 42±4 minutes into nc14, at which the information caried by the gap genes is maximized [11]. As noted above, this time window precedes our *eve* data by ~8 min, accounting for the expected delay [18] for the impact of changes in gap genes concentrations on Eve protein levels.

Given the precise positioning of Eve stripes in gap mutants and considering that during nc14 the gap genes are providing the main input source for *eve* expression (within the embryo's main trunk) [4], we ask whether a corresponding level of reproducibility is found at the level of gap gene concentrations. It is traditionally expected that spatiotemporal expression profiles in mutants are more variable as the animals are assumed to be less capable in coping with external stress. In our dataset the opposite is observed: although there are significant deviations for the mean expression profiles, in all mutant backgrounds, expression levels are highly reproducible from embryo to embryo (Figure S2D). Variances around the average profile, even when that profile is considerably displaced compared to wild-type, are close to the levels observed in wild-type (Figures S2D–F). While the positional error measured for hemizygous mutants is as accurate as the one for wild-type (i.e., ~ 1% EL, Figure S2E), homozygous gap mutants show slight increases in the positional error obtained from the remaining 3 genes (Figure S2F). This increased variability typically appears in the boundary regions of the removed gene. While the positional error can locally go as high as ~6% EL (e.g. in 0X-*Kr* around the region of reversed polarity duplications), at Eve stripes positions it typically reaches around 2% EL, matching the upper bound of the variability measured for the homozygous eve stripes positions (Figure 1I–J).

**Gap gene combinations in wild-type embryos and *eve* precision**

To investigate the relationship between the gap gene patterns and *eve* expression, we used our four independent measurements in 2X embryos to extract the wild-type gap gene concentrations present during the eight minutes immediately preceding the time window for which *eve* expression is measured. By examining locations corresponding to peak and trough locations along the AP-axis, we established the gap gene expression range within which *eve* activity is fully turned on or turned off, respectively. For each wild-type Eve stripe we determine its average peak position along the AP axis, which we know to be reproducible to within ~1%, and then extract at that location in each 2X embryo the quadruplet of simultaneously labeled gap genes expression levels. While the concentration for individual gap genes in these embryos at any point along the AP axis can vary by as much as 25% of their average wild-type levels (Figure 3A & S3A), for each position along the AP axis, our analysis identifies at least one gap gene whose level is sufficient to provide a positional error that is close to the 1–2% EL level (Figure 3B).

Curiously, the dynamic range of activity ($2\sigma_i$ around the mean activity) for some specific high-precision-providing gap genes is large (Figure 3C and S3C). However, the positions at which they provide such high precision coincides with the gap gene pattern boundaries. Thus, the large dynamic range in activity is offset by a large local derivative (dg/dx), considering that the positional error $\sigma_x$ is given by the ratio of $\sigma_i$ and dg/dx. This argument is supported by a strong correlation between $\sigma_i$ and the local derivative measured for the gap gene providing the highest accuracy level (Figure 3D). Therefore, cases where gap genes with wider dynamic ranges are providing the observed accuracy levels are enabled by a steep change of expression levels between adjacent cells. This image sits well with the increased positional error in homozygous mutants around these boundary regions of the removed gene (Figure S2B). While single gap genes might therefore account for the precision observed in *eve* expression, other gap genes are not irrelevant - their levels may serve to distinguish different broad regions of the embryo rather than defining a precise cell identity [4].

To investigate this phenomenon further, we performed a PCA analysis on all wild-type combinations found at eve peaks. The values of the first two principal components (i.e., the ones containing most of the variance) when exposed in a scatter plot (Figure 3E and S3E) are perfectly disjoint for all Eve peaks. In addition, troughs fall in the interstitial space mapped out by the peaks, providing an intuitive picture for single cell accuracy of the system: for a given pair-rule gene there are exactly two cells positioned between an adjacent peak-through pair. Overall, these results demonstrate that a large concentration range in the code words repertoire can coexist with the observed precision in the downstream gene expression profiles.

**Code word repertoire in hemizygous gap mutants**

The quadruplet of gap genes associated with a given expression domain of the Eve pattern can be thought of as a stripe specific *code word* that elicits *eve* activity. To probe the locus' response to our genetic manipulations, we examine the resulting code word perturbations in the altered genetic backgrounds. In this section, we focus on the hemizygous mutants (1X). While Eve stripes may be slightly displaced, all four of these 1X backgrounds display seven stripes with 100% penetrance, and each stripe can be unequivocally matched to its corresponding stripe in wild-type. We can thus measure code words in the 1X embryos at the locations of the 1X mutant stripes, and compare them to the respective 2X counterparts, measured at wild-type Eve stripe locations.

Of the four concentrations composing a code word, one corresponds to the manipulated gap gene. While it is nominally (i.e., genetically) halved in the 1X case, the actual average concentration measurement across embryos does not always show a 50% reduction from the corresponding wild-type level (i.e. y=0.5x line in Figures 4A and S4A-D). This observation is potentially due to compensatory effects within the gap gene network. We further find that all other mean data points, stemming from the three remaining concentrations of each code word corresponding to the unmanipulated gap genes, are unchanged within error bars from their corresponding wild-type levels (i.e. diagonal, y=x, in Figures 4A and S4A-D). Hence Eve-stripes in 1X gap gene backgrounds have one concentration almost halved in their code word, while the other three concentrations are unchanged and match the wild-type levels (Figure 4A). The Eve locus thus displays a tolerance to 50% dosage perturbation for one gap gene, and otherwise activates at wild-type concentration levels for the remaining three gap genes. Note that gap gene expression levels here are compared between the mutant stripe and the wildtype stripe positions. Since stripe positions in the hemizygous mutant can often shift (Fig1 E&G and 1S B), the comparison of levels in these cases is between different positions along the 2 "averaged embryos".

However, the preservation of the unperturbed gap genes levels responsible for activating the mutant *eve* peaks is not indicative of a lack of change in the activity of these genes. We observe clear concentration changes between wild-type and mutant conditions when comparing their levels at *identical* positions along the AP-axis (Figure 2F–I). In cases where preserved code word levels are accompanied by anterior shifts of the corresponding Eve stripes (green data points in Figure 4A), these shifts are associated with subtle but significant concentration adjustments in the levels of other gap genes, compared with gap levels at the same position in the wildtype -- where no stripe normally appears. These gap genes are thereby 'compensating' their level in response to the direct effect of the external manipulation, and the eve locus recognizes and responds to the familiar code words in their shifted positions. Indeed, the increase in concentration of one of the unperturbed genes correlates proportionally with the magnitude of the positional shift of the corresponding Eve stripe (Figure 4B). Thus, the genetic perturbation in one gene can induce a change in the concentrations of the other genes.

As an example, let's examine Kni concentration changes in a *1X-Kr* mutant and their effects on Eve stripes 4 and 5 that overlap with both the Kr and Kni domains (Figure 4C). Both stripes display a significant anterior shift. Stripe 4 occurs where Kr concentration is halved compared to wild-type. Yet the shifts of stripe 4, and more pronouncedly that of stripe 5, follows the shifted pattern of Kni in the 1X-*Kr* background, such that Kni concentration is identical to that found in these stripes under 2X-*Kr* background (Figure 4D,E). Note that *hb* and *gt* are barely expressed here, and hence their concentration in the 1X-*Kr* mutant are effectively unchanged from wild-type, merely completing the code word. Therefore, a single unperturbed gene accounts for the position shift (Figure 4B); Eve is activated at the position that can accommodate the imposed halving concentration of one gene with the least amount of further adjustments. These results are recapitulated in our PCA when overlaying the 1x code words with the wild-type code words (Figures 4F). While the dynamic ranges over which locus responses are elicited increase slightly, the averages remain similar and peaks and troughs are still precisely separable (Figure S4G).

This correction capacity is not limited to code words associated with Eve peaks: while the spatial extent of the gap gene response is wide in the hemizygous mutants, gap code words along the entire main trunk fall *within* the wild-type repertoire. The active response of the other gap genes is wider than that spanned by the direct perturbation (i.e. the wild-type domain of the halved dose gene, Figure S4E). A $\chi^2$-analysis quantifies the similarity between mutant code words and the distributions of wild-type code words, at any position along the body axis (Figure S4F). Code words responsible for *eve* expression in hemizygous gap mutants are mostly contained within the wild-type repertoire, with an overlap of ~99.9% between the two $\chi^2$-distributions -- despite the relatively wide reach of deviations from wild-type concentrations at a given position. Together these results point to a strong tendency of the gap gene network response to actively compose familiar concentrations combinations upon the removal of a single copy of a gap gene.

**Homozygous gap mutants mostly display canonical code words**

Interestingly, the $\chi^2$-analysis of the homozygous null mutants (0X) only shows radically new combinations for ~10% of the code words (Figure S5A). Among the total of 24 Eve stripes detected in our 0X data sets we only detect three Eve stripes where large deviations in concentration profiles always suggest novel code words. Even under a more permissive threshold for novel code words, the distance from the wild-type manifold only increases beyond its baseline for code words in regions of the embryo overlapping with the wild-type expression domain of the manipulated gap gene. For the homozygous null mutants too, the spatial range of significant response always extends beyond the removed gene domain (Figure S5B). In all gap mutants, the code words that result in *eve* expression cluster around the familiar code words despite the genetic manipulation. This suggests an active tendency of the network to achieve or approximate the code word repertoire present somewhere along the AP axis of the wild-type embryo. This effort is obviously more successful where the perturbations are relatively small.

Given that most code words are canonical, we can rely on similarity to wild-type to infer stripe identity in the homozygous null mutants. To better understand the origin of these stripes, we resort back to our principal component analysis and consider the first two principal components of the code words for the four cases of the null mutant backgrounds (Figure 5A–5D). In the reminder of this section, we integrate our PCA interpretation for the homozygous null mutants with reporter assays for minimal stripe enhancers of *eve*.

In the case of a complete removal of *Kr*, five stripes (*I–V*) are observed (for clarity we denote mutant stipes with roman numbers sequentially from anterior to posterior). The identity of these stripes can be easily read off the PC1 vs. PC2 plane: while I/II and IV/V seem to largely recapitulate the anteriorly-most and posteriorly-most wild-type stripes, respectively, stripe III seems to be a duplication of wild-type stripe 7, right in the region where the central *Kr* domain would be. This view is consistent with the polarity duplication observed in the final cuticle of *Kr* homozygous embryos [21,23], although the identity of the associated abdominal segment was unclear. Even though all code words in *0x-Kr* remain within the wild-type repertoire (according to the $\chi^2$-analysis), reporter assays indicate simultaneous activity of both the Eve stripe 3 and 7 enhancer [14] as well as the Eve stripe 2 enhancer, which extend posteriorly towards the position of stripe III [16]. This joint contribution matches well with the proximity of stripe III code words in our PCA analysis to both stripe 7 and stripe 2. Furthermore, stripe II shows relatively low penetrance (69%), which is echoed by several code words that are falling out of the wild-type repertoire for stripe 2.

Replicating this analysis for the *0x-kni* mutants, for which six stripes are observed, stripes I/II/III and VI seem to replicate the anteriorly-most three and the most posteriorly wild-type stripes, respectively (Figure 5B). Mutant stripe IV overlaps with wild-type stripe 3, along its border with stripe 4, matching reporter activity of both minimal stripe enhancer 4/6 [13] and 3/7 [14] at this position. Stripe V code words not only activate both enhancer elements 4/6 [13] and 3&7 [14], but there is even some activation from 1 and 5 enhancers [13]. They match the occupancy of the corresponding code words in the empty region between stripe 7 (best match) and stripes 4 to 6. While cuticle interpretations for this mutant suggested a merger of stripes 3 to 7, the actual overlaps of stripe IV with 3 and stripe V with 7 in our PCA analysis suggest that the region between stripe III and VI is not simply merging stripes 3 to 7, but at least contains duplications of stripe 3 and 7. Stripe V also show 58% penetrance, with some embryos composing novel code words according to the $\chi^2$-analysis. Nonetheless, the *eve* locus is continuously active above baseline at this defuse expression region, for all *0x-kni* embryos.

In *0x-gt*, according to the $\chi^2$-analysis, stripe I is activated by entirely novel code words (Figure 5SX), however in the PCA analysis the corresponding data points largely overlap with the wild-type stripe 2 code words (Figure 5D). According to reporter literature, stripes 1 [13] and 2 [15] are both active at this position. Stripes II, III, IV, V, and VII match stripes 2, 3, 4, 5, and 7, respectively. Stripe VI is positioned in the region between the wild-type stripes 4 ,5 and 6, matching reporter experiments where both stripe elements 1/5 and 4/6 are contributing to *eve* expression at this position [13]. This stripe falls within the wild-type manifold with elevated $\chi^2$ levels (Figure S5X) and show partial penetrance (89%).

Lastly, in *0x-hb* stripes I and II are both activated by novel code words and show partial penetrance (79% and 50%, respectively), with stripe I most similar to stripes 6 and 7 and II most similar to stripe 4. While weak activation of reporter 1/5 [13] is detectable at stripe I position (also proximal to stripe 7 in the PCA), the similarities to 4, 6 and 7 are compatible with overlapping anterior activity of *eve* reporters 4/6 [13] and 3/7 [14]. These posterior identities of the anterior mutant stripes are clearly not aligned with those cells' final fates, as these embryos form regular head structures. However, it does match the duplication and reverse polarity observed in homozygous *hb* null mutants when both zygotic and maternal contribution are removed. It is unclear how such a 'rescue' from the TFs' prediction is achieved by maternal Hb. Stripes III, IV and V match stripes 4, 5 and 6, respectively, whereas stripe VI is positioned between stripe 6 and 7. The posteriorly expanded activity of 4/6 ([13]) and anteriorly expanded 3/7 activity ([14]) results in a quasi-merging of stipe VI with both stripes 6&7.

**Discussion**

A code word can be thought of as a functional combination of transcription factors that is read by a gene locus and participates in defining a cell's fate. In the early fly embryo, the gap genes compose code words that are read by the Eve locus, and entirely determine its transcriptional output in the main trunk [4]. Here we show that a wide repertoire of gap gene code words activates the Eve locus in the wild-type, and that an even wider repertoire is functional under the genetic background of gap mutants. Nevertheless, in both wild-type and gap mutants, the accuracy of stripe positioning remains near ~1%, compatible with a single cell precision. For the homozygous nulls, positional accuracy of eve stripes is accompanied by partial penetrance of certain stripes, accounting for their renown variable body plan across embryos of an identical genetic background [22]. Code words in gap mutants tend to show low distance from the wild-type repertoire, at least in regions of the embryo outside the wild-type expression domain of the perturbed gene, and often within it. This conserved accuracy and familiarity of code words is a result of active efforts of the system: nuclei responding to the genetic manipulation cover a wider spatial range than the directly affected gene. This suggests that the gap genes network is providing robustness against possible disturbances and bear the ability to adapt its expression under ongoingly changing developmental environment.

<u>Definition of non-canonical code words</u>

Throughout this manuscript we used 2 different measures to assess similarity between wild-type and mutant code words: $\chi^2$-analysis (Figure S5); and PCA (Figure 4&5), by proximity of mutant code words to the wild-type repertoire in the $PC1^-$ $PC2$ plane. These measures are not interchangeable for identifying similarity between code words, and each comes with its own set of limitations. While according to the $\chi^2$-analysis nearly all code words in hemizygous mutants are contained within the wild-type repertoire, deviations from the wild-type repertoire are identified for these mutants by the PCA, where hemizygous code words occupy a notably wider regions of the PC1&PC2 plane. According to the $\chi^2$-analysis, increased distance is observed (nearly) only for homozygous mutants, in embryo regions overlapping with the perturbed wild-type domain. In addition, deviations of homozygous mutant code words seem more extreme according to the $\chi^2$-analysis, compared with their deviations in the PCA. It is also important to consider that the PCA is using partial information, whereas the $\chi^2$-analysis is using the full information comprised by all four gap genes; however, as any defined metric, it is also biased by dictating a certain geometry for how the distance from the wild-type repertoire should be measured.

According to the PCA, the vast majority of mutant code words fall close to the wild-type repertoire, rather than occupying extremely distant regions. In this case, determining whether a code word is a novel one or not matters mostly via its function: whether its sensors – the transcriptional machinery -- can read it and produce a functional output. While both canonical and non-canonical code words activate the eve locus above its baseline, resorting to reporter assays in the homozygous gap mutants shows that code-words in domains where the deleted gap gene would normally have been expressed drive *eve* expression contributions from more than one "stripe-specific" enhancer [13–16]. Since this loss of stripe specificity is never observed in wild-type embryos, such non-canonical activation may truly mark non-canonical code words.

<u>Molecular corollary of code words and spatial specificity of locus activation</u>

Can code words be translated into enhancer activities? One property of enhancers is their specificity in binding of TFs combinations, e.g., the concentration sensitivity of each Eve stripe enhancer. Integrating data from reporter assays of enhancer specific activity under homozygous null background with our PCA reveals a remarkable accordance between the proximity in the PC1 and PC2 plane and the activation of minimal stripe enhancer elements [13–16]. It is possible that the loss of specificity underlies the variable body plan of homozygous gap mutants: the partial penetrance of stripes in the homozygous mutants could potentially stem from different efficacies of activation of the multiple enhancers involved. Small deviations of the concentrations found in individual embryos could lay near the activation threshold of an enhancer whose specific code word is more distal in the PCA.

Do the concentration dependent behaviors implied in our definition of code words provide a sufficient picture of spatial activation of Eve? Although our measurements are consistent with earlier reporter studies, it is possible that specific combinations of gap genes may determine activity only under the exogenous conditions used in most reporter studies. Such reporter-based assays may not duplicate the more complicated interaction with the rest of the genome that may be present in the endogenous control regions [24,25]. Therefore, observations obtained with exogenous reporters may reflect a more direct concentration dependence of the enhancer sequences themselves. It is possible that under endogenous conditions, the dynamic tertiary structure of the DNA around the eve locus is adding a control layer that enables specificity of activation even in gap mutants. Measuring endogenously labeled activity under mutant background is required to determine whether such non-canonical code words do indeed induce non-specific activation. Such experiments would teach us more about the scope of contribution of code words to transcriptional control and whether multi-activation of enhancers is an artificial result of the exogenous context.

Network response to genetic perturbations and the role of zygotic genes

Our analysis provides a straightforward intuition for the origin of the mutant Eve stripes. Together, the integrated interpretation of our PCA and the accumulated reporter assays portrays an image where multiple *eve* enhancer elements are simultaneously active, in accordance with their proximity in the PC1–PC2 plane. Moreover, non-specific activation of *eve* reporters occurs in transmuted positions overlapping with the removed wild-type domain, where often elevated distances from the wild-type repertoire is registered and partial stripe penetrance is observed.

By and large, the gap network shows a strong tendency to compose familiar code words. Even when non-canonical gap combinations activate multiple position-specific enhancers at a given nucleus, the activation pattern is following similarity to canonical code words normally activating each of these position-specific enhancers. The combination of similarity to canonical code words and the loss of position specificity marks the limit of the functional flexibility of these enhancers. As with the hemizygous mutants, this adherence to the wild-type repertoire stems from active work invested by the gap network in response to the removed gene (Figure S5C). Such correction capacity of gene networks may underly the robustness manifested by the developing fly to a variety of documented perturbations [26–28].

The ability of the system to correct for small perturbations is apparent for all gap mutants, both hemizygous and homozygous. This correction capacity seems more limited upon stronger perturbations. Yet, even when a gap gene is completely removed, the combinations of gap genes that are found in the mutant embryo are mostly familiar (canonical) code words, including in regions that have clearly undertaken an active response to the imposed genetic perturbation. This capacity of the system to cope with perturbations and adjust, potentially contributes to the reproducibility and precision observed in the system [11,29]. Since in the fly embryo most of the positional information is already determined by maternal morphogens [4,30], our data suggest that a pivotal role of the zygotic component of the segmentation gene network is to ensure accuracy and precision under unexpected disturbances to the developmental process.

## Methods

**Key resources table**

| REAGENT or RESOURCE | SOURCE | IDENTIFIER |
|---|---|---|
| Experimental models: Organisms/strains | | |
| D. melanogaster: Oregon-R, wild-type | Lab stock | Flybase: FBst1000077 |
| D. melanogaster: hb mutation | Eric F. Wieschaus (Princeton) | N/A |
| D. melanogaster: Kr mutation | Eric F. Wieschaus (Princeton) | N/A |
| D. melanogaster: kni mutation | Thomas Gregor (Princeton) (ML CRISPER) | N/A |
| D. melanogaster: gt mutation | Thomas Gregor (Princeton) (ML CRISPER) | N/A |
| Antibodies | | |
| Gap staining antibodies (guinea pig anti-Gt) | Eric F. Wieschaus (Princeton) | N/A |
| Gap staining antibodies (rat anti-Kni) | Eric F. Wieschaus (Princeton) | N/A |
| Gap staining antibodies (rabbit anti-Kr) | Eric F. Wieschaus (Princeton) | N/A |
| Gap staining antibodies (mouse anti-Hb) | Eric F. Wieschaus (Princeton) | N/A |
| Secondary antibodies Gap staining: Alexa-514 (rabbit) | Invitrogen, Grand Island, NY | Cat# A31558 |
| Secondary antibodies Gap staining: Alexa-568 (guinea pig) | Invitrogen, Grand Island, NY | Cat# A11075 |
| Secondary antibodies Gap staining: Alexa-647 (Rat) | Invitrogen, Grand Island, NY | Cat # A21247 |
| Secondary antibodies Gap staining: Alexa-430 (mouse) | Invitrogen, Grand Island, NY | Catalog # A-11063 |
| Eve staining antibodies (mouse anti-Eve) | Eric F. Wieschaus (Princeton) | N/A |
| Eve staining antibodies (guinea pig anti-Gt) | Eric F. Wieschaus (Princeton) | N/A |
| Eve staining antibodies (guinea pig anti-Kr) | Eric F. Wieschaus (Princeton) | N/A |
| Eve staining antibodies (guinea pig anti-Hb) | Eric F. Wieschaus (Princeton) | N/A |
| Eve staining antibodies (guinea pig anti-Kni) | Eric F. Wieschaus (Princeton) | N/A |
| Secondary antibodies Eve staining: Alexa-647 (mouse) | Invitrogen, Grand Island, NY | Cat# PIA32728 |

| Secondary antibodies Eve staining: Alexa-568 (guinea pig) | Invitrogen, Grand Island, NY | Cat# A11075 |
|---|---|---|
| Software and algorithms | | |
| Custom MATLAB code | This paper | N/A |

**CONTACT FOR REAGENT AND RESOURCE SHARING**

Further information and requests for resources and reagents should be directed to and will be fulfilled by the Lead Contact, Thomas Gregor (tg2@princeton.edu).

**EXPERIMENTAL MODEL AND SUBJECT DETAILS**

*Fly strains*

All the embryos from a cross between a pair of hemizygous parents for one of the 4 gap genes were obtained by allowing them to lay eggs for 2 hours, and then maturing the eggs for 2 more hours. Stocks were balanced as follows:

cn bw $Kr^1$ / SM1

$hb^{12}$ st e / TM3, Sb hb lacZ "kni-null" allele - a CRISPR-mediated replacement of the kni region (upstream regulatory regions and coding region) with a 2attp-dsRed cassette was performed. The homology arms were amplified from the genomic DNA of the nos-Cas9/CyO injection line (BDSC #78781). The two Cas9 cutting guide RNAs sequences used are [GGGAGGGCTTGATTCGGGAAAGG] and [CTTGAAGCTCATTAATTCCACGG]. PCRs from the dsRed to the flanking genomic regions were performed to verify the deletion. The allele was balanced with TM3, Sb balancer.

"gt-null" allele was produced in a similar manner. The cas9 injection line used was BDSC #51324. The two Cas9 cutting guide RNAs sequences used are [CGGCCGGCGAGGAAGTGAACGGG] and [TCTTACGTGTAAGAATTCATGGG]. The allele was balanced with FM7 balancer.

**METHOD DETAILS**

*Measuring gap gene expression*

Protein levels of Gap genes were measured as previously reported [17] with slight adjustments as follows: we used rabbit anti-Kr along with mouse anti-Hb, guinea pig anti-Gt and rat anti-Kni. Secondary antibodies are, respectively, conjugated with Alexa-430 (mouse), Alexa-514 (rabbit), Alexa-568 (guinea pig), and Alexa-647 (rat) from Invitrogen, Grand Island, NY. Expression levels were normalized such that the mean expression levels of WT embryos ranged between 0 and 1, with background subtracted from individual embryos (the minimal value measured along its dorsal AP profile) and divided by the maximal value of the mean wild-type dorsal profile measured per gene. Specifically, gene expression profile $I_\alpha^g$ of an individual embryo $\alpha$ of any genotype was calculated as:

$$I_\alpha^g = \frac{I_{\alpha\,raw}^g - \min|_x(I_{\alpha\,raw}^g)}{\max|_x(<I_{wt}^g>)}$$

Where $\min|_x(I_{\alpha\,raw}^g)$ is the lowest raw fluorescence intensity of embryo $\alpha$, and $\max|_x(<I_{wt}^g>)$ is the highest raw fluorescence intensity value of the mean wild-type embryo fluorescence profile; $I_{\alpha\,raw}^g$ is the raw fluorescence profile of an individual embryo of any genotype (mutant or wild-type).

*Measuring eve gene expression*

To image Eve, we used mouse anti-Eve together with a guinea pig anti-Gap (Hb/Gt/Kr/Kni), according to the mutant stock we were imaging, to sort out genotypes of individual embryos. Secondary antibodies are, respectively, conjugated with Alexa-647 (mouse), and Alexa-568 (guinea pig) from Invitrogen, Grand Island, NY. Embryo fixation, antibody staining, imaging and profile extraction were performed as previously described [4,17]. We used *eve* expression to extract the locations of expression troughs and peaks. Eve protein profiles were simultaneously measured in mutant and wild-type embryos in the time widows of 45- to 55-min into n.c. 14. Expression profiles were normalized as with the Gap staining, such that the mean expression levels for each gene in the wild-type subpopulation of embryos in each measurement ranged between 0 and 1, as described above for the gap gene measurement.

*Quantitative comparison of gap protein levels across genotypes*

As before [4], expression levels in mutants were measured quantitatively to enable comparison of their levels to those found in wild-type. To that end, homozygous and hemizygous mutants as well as wild-type embryos obtained from each hemizygous gap stock were fixed together, stained together, mounted together in a random order on a single slide and then imaged sequentially in a single session. As indicated above, our normalization of the fluorescence signals of gap genes from all mutant embryos results in comparable wild-type units for each gap gene across hemizygous, homozygous and their internal control of balancer over balancer (2xGap).

*Genotype and time-stamp assignments*

In order to sort out genotypes, we first assigned each embryo with its time stamp estimation according to the progression of cellularization front [17,31]. Next, we pooled embryos from a short time window together and segregated the hemizygous null embryos by threshold, since they were all naturally separable, and used automatic Kmean clustering, with k=2 for separating the hemizygous from the wild-type (balancer over balancer). We repeated this process with overlapping time windows, such that every embryo was assigned with a genotype 5 times, and the most likely genotype was selected. Embryos that showed ambiguous genotype according to repeated assignments were excluded. The resulting genotype proportion largely matches the mendelian expectation under the genetic conditions involved (25% wt, 25% homozygous null and 50% hemizygous mutants, except for *1xgt*x*1xgt* resulting in 50% wt and 25% of each hemizygous and homozygous mutants due to dosage compensation of the Gt protein in males, since *gt* resides on the X chromosome).

**Quantification and Statistical Analysis**

*Identifying eve peaks and troughs*

Peak positions were programmatically identified per embryo as local maximum expression for a maximum of N peaks along the main trunk, with $N_{peaks} = 7$ for all wild-type and hemizygous mutants, and $N_{peaks} = [6,5,7,6]$ for 0xhb, 0xKr, 0xgt and 0xkni, respectively. To avoid identification of small fluctuations of expression, a minimal inter-peak-interval was defined as 3-4 cells. Additionally, hemizygous nulls profiles were less smooth, which often resulted in splitting a single clear major peak to 2 or more adjacent sub-peaks and were therefor slightly smoothed (by 2% egg length) to best identify the location of maximal expression. To identify eve troughs, the profiles were inverted, and an identical process was performed for $N_{troughs} = N_{peaks} - 1$. For the hemizygous null data, a minimal peak prominence was also defined to avoid identification of very small fluctuations in expression as troughs.

Since in all the homozygous mutants the number of stripes and their locations varied across the population of embryos, the assignment of the serial number of a null stripe was performed as follows: after peak identification, the subset of embryos exhibiting the maximal number of stripes were pooled, and the mean position of each null stripe was calculated from this subset. The remaining embryos showing a partial set of stripes were assigned with a serial number per stripe according to its minimal distance from all the possible stripes in that genotype.

*Estimating positional error and position displacement of eve stripes*

Positional error of eve is defined as the standard deviation of eve peak (or troughs) locations across the population of embryos of the same genotype. The displacement was defined only for the mutant embryos, as the distance between the stripe position of an individual mutant embryo from the mean position of the same stripe in the mutant. Therefor the error of the displacement of a given stripe is by definition the positional error of that stripe. Since for the nulls stripes identity was unclear, Figure 1 and Figure 1S display the minimal displacement from all stripes.

Positional error measured at the gap level was performed for a single gene, 3 or 4 genes as previously described [11,12].

*Estimating deviations from wild-type levels of gap proteins along time*

To estimate the deviations from wild-type expression in mutant expression along time and across the egg we reconstructed the mean expression for each genotype using a sliding window of 8 minutes, such that the mean expression at any minute along nuclear cycle 14 was computed from embryos with time stamps of ±4 minutes around that time. The earliest time point for which this mean level was reliably estimated for all genotypes was 10 minutes, and due to the nature of the sliding window, the latest time point in the time series is 56 minutes (including embryos from 52-60 minutes into nc 14). After reconstructing this time series for each of the genotypes, the mean levels of the wild-type were subtracted from the mean levels of the mutant, such that positive differences indicate overexpression in the mutants (red shades in the kymographs of Figure 2), and negative differences indicate reduced expression of the mutants (blue shades) compared with the wild-type level at any position along time:

$$\Delta I^{gene}_{genotype}(x,t) = \Delta I^{gene}_{genotype}(x,t) - \Delta I^{gene}_{wildtype}(x,t)$$

To estimate the noise level of expression intensity in our measurement, we bootstrapped the wild-type time series (n=400 repeats) and computed the difference between pairs of its resampled versions. Examples for these 200 resampled kymographs for the 2xGap are displayed at the leftmost column of Figure 2 for each measurement of a single gap gene manipulation. The noise level for each measurement session was estimated as ±2 standard deviations of the resampled difference of 2xGap per kymograph for all its position and times. The resulting difference, $\Delta I^{gene}_{genotype}(x,t)$, below that level is below the limit of our measurement accuracy (demonstrated as grayed out pixels in Fig S2). The maximal of these standard deviations across all 16 kymographs for the 2xGap is 0.04 of the wild-type maximal expression per gene. We therefor use 0.08 (±2σ) as the noise level limit when comparing the mean level found in wild-type and the mutants (grayed area around the unity line in Fig 4 A,D &E, and in Fig S4 A-D). This procedure was only computed for the gap data and not the eve data, since in this work we only used eve data for its positional ques and not its intensity.

*PCA of the wild-type repertoire of eve activation code words*

PCA was performed on gap quadruplets from individual wild-type embryos with time-stamps between 38-46 minutes into NC 14. In Figure 3 the PCA is performed on the pooled wild-type embryos from all 2Xgap datasets, whereas in Figure 5 the PCA is performed on wild-type embryos from each internal control separately. In Fig S3 the PCA is performed on wild-type embryos from 6-60 minutes into NC 14. To explore the similarity between the wild-type and the mutant gap code words that activate eve stripes, we projected the gap quadruplets from any other genotype onto the first 2 PCs obtained from the wild-type.

$\chi^2$ *—analysis*

To estimate the novelty of gap code words that activate eve stripes in the gap mutants we resorted to the $\chi^2$ per-gene measure [4], used to estimate the distance of a gap quadruplet from the distribution of gap quadruplets found at a given position along the egg:

$$\chi^2_K/K = (\{g_i\}, x) = \sum_{i,j=1}^{K} (g_i - \bar{g}_i(x))\left(\hat{C}^{-1}(x)\right)_{ij}(g_j - \bar{g}_j(x)))$$

With K =4 the number of genes used, $\{g_i\} = \{g_1, g_2, g_3, g_4\}$ are Hb, Kr, Gt and Kni levels at a given position x, $\bar{g}_i(x)$ is the respective mean level across embryos, and $\hat{C}$ is the 4X4 covariance matrix of $\{g_i\}$. Since estimating the covariance is extremely sensitive to the measurement noise, we made sure that our gap data recaptures the 1% accuracy levels previously reported [4,11,12], as can be seen according to the $\sigma_x 4g$ for our internal control within the main trunk (Figure S2).

For the mutants, we took the minimal $\chi^2$ per-gene compared with the wild-type reference distributions at any position along the egg. Where this distance exceeded the maximal distance found for a wild-type code word from the distribution of code words found at its position across a population of wild-type embryos, the mutant code word is marked as non-canonical, or novel to the wild-type repertoire. Even regardless of this distance cutoff, the $\chi^2$ per-gene deviates from its baseline mainly for the homozygous null mutants, and only regions of the embryo that overlap with the wild-type expression domain of the removed gene. Note that for Kr nulls the correction capacity of the gap network is near perfect such that at any position along the egg canonical code words are composed, even at the absence of Kr.

*Average nuclear response to genetic manipulation*

In order to estimate the cellular response to the genetic manipulation in stake, we used the deviation from the wild-type expression, $\Delta I^{gene}_{genotype}(x,t)$, and took the square-root of its average squared values over all positions and times, such that both increased expression and decreased expression relative to the wild-type receive equal weight, and each kymograph gets an estimate of response intensity within NC 14:

$$R^{gene}_{genotype} = \sqrt[2]{\frac{\sum_n (\Delta I^{gene}_{genotype}(x,t))^2}{n}}$$

Where n is the number of pixels over all times and position per kymograph (n=800*47). To control for this value from each mutant kymograph, it is compared with $R^{gene}_{genotype}$ obtained from all the bootstrapped versions of its internal control of 2xGap (n=200*800*47).

For each Kymograph we also calculated the spatial response by taking the square-root of the averaged square deviation across time, for each position:

$$R^{gene}_{genotype}(x) = \sqrt[2]{\frac{\sum_t (\Delta I^{gene}_{genotype}(x,t))^2}{t}}$$

Similarly, here we compare the resulting value per position and gene with that obtained from all the bootstrapped versions of its internal control (n=47*200).

**QUANTIFICATION AND STATISTICAL ANALYSIS**

Gap gene protein expression in mutant backgrounds alongside with their internal control (*2xgap*) was measured throughout nuclear cycle 14. Embryos from a given hemizygous-null gap stock were simultaneously fixed and stained for the 4 trunk gap gene proteins, such that all 3 genotypes from a given hemizygous gap gene stock were imaged sequentially in a single session. We imaged n = 698 embryos from the 1xhb stock, out of which 140 are identified as 2xhb embryos, 365 as hemizygous to *hb* null (*1xhb*) and 193 as homozygous to *hb* null (*0xhb*); n = 1189 embryos from the 1xgt stock, out of which 608 are identified as *2xgt* embryos, 296 are *1xgt*, and 285 are *0xgt*; n = 703 embryos from the *1xkni* stock, out of which 178 are identified as *2xkni* embryos, 352 are *1xkni*, and 173 *0xkni*; and n = 496 embryos from the *1xKr* stock, out of which 94 are identified as *2xKr* embryos, 241 as *1xKr*, as 134 *0xKr*.

Eve protein levels were simultaneously labeled along with the gap gene under manipulation per stock, to enable genotype identification. Embryos from a given hemizygous-null gap stock were fixed, stained and imaged sequentially, focusing on embryos within the time window of 45-55 minutes into nc 14. We imaged n = 53 embryos from the *1xhb* stock, out of which 12 are identified as *2xhb,* 27 as *1xhb* and 14 as *0xhb;* n = 77 embryos from the *1xgt* stock, out of which 40 are identified as *2xgt,* 18 as *1xgt* and 19 as *0xgt;* n = 71 embryos from the *1xkni* stock, out of which 18 are identified as *2xkni,* 34 as *1xkni* and 19 as *0xkni;* and n = 107 embryos from the *1xKr* stock, out of which 23 are identified as *2xKr,* 49 as *1xKr* and 35 as *0xKr.*

# Figure legends

**Figure 1: Precise patterning in gap gene mutants, with partial penetrance for stripes in homozygotes.**

(A) In the early *Drosophila* embryos four gap genes control the expression pattern of the *eve* locus. Cartoon shows the spatial order of gap gene patterns (top) and the seven Eve stripes (bottom). Color codes represent the different gap genes and the identities of the Eve stripes. (B) Genetic design of gap gene manipulations. For each of the major four gap genes hemizygous parents with a null mutation in one gap gene on one chromosome generated offspring with 50% hemizygotes (1X), 25% homozygotes (0X), and 25% quasi wild-type (2X, i.e. two non-viable balancer chromosomes). (C) Positional displacement (Δx, left) versus positional error ($\delta x$, right) shown as a cartoon. Displacements correspond to mean position shifts of a pattern; error corresponds to fluctuations of a given pattern boundary around the mean. (D) Eve patterns for 12 embryos are shown at 50±5 min into nc14 for the 2X-*hb* experiment (gray). Mean Eve pattern (green); automated identification of peak locations for seven stripes (circles); mean Hb pattern (dashed gray). In bottom panel, error bars represent the positional error ($\delta x$) for each Eve stripe. Horizontal dashed gray lines represent ±1% EL. (E) Same as in D for hemizygous 1X-*hb* experiment. Eve patterns shown for 27 embryos, mean pattern in yellow, with 2X pattern overlayed in green. Bottom panel shows mean positional displacements for each Eve stripe, with positional error indicated by error bars. (F) same as in E for 0X-*hb* experiment; 14 Eve patterns (gray) with mean in magenta. Stripe penetrance is annotated if different from 100%. (G) The distributions for Δx across all homozygous (magenta) and hemizygous (yellow) embryos (n=215, for a total of 51 stripes). (H) Stripe penetrance homozygous null mutants (by gap color code), indicating the percentage of embryos showing a certain number of stripes (n=24 stripes in n=87 embryos). The distributions of $\delta x$ in each genotype is presented for Eve peak (I) and trough (J) locations using genotype-specific color code. See Figure S1 for Eve patterns in all genotypes.

**Figure S1: Precise Eve patterning in gap gene mutants.**

A total of 12 experimental conditions is analyzed: 2X, 1X, and 0X genotypes for four major gap gene crosses (Figure 1A). (A) Eve patterns for four 2X experiments with embryos fixed at t=50±5 minutes into nc14: individual patterns (solid gray lines), mean (green), and peak positions (circles). The mean wild-type expression profile of the perturbed gap gene is shown in dashed gray. Bottom panels show stripe displacement (Δx, zero by definition for 2X embryos) and positional error ($\delta x$, error bar) for each Eve peak. Horizontal dashed gray lines represent ±1% EL. (B) Same as in A for all 1X experiments. (C) Same as in A for all 0X experiments; since stripe identity is unknown in 0X case, only the minimal displacement (Δx, bottom insets) from any stripe is presented. Some stripes only emerge in a subset of the embryos, in these cases, the percentage of embryos presenting a stripe is annotated (bottom insets). (D-F) Examples for Eve expression profiles from three 0X-*hb* embryos where different stripe combinations can be observed. Stripe serial number is determined automatically, based on minimal displacement from any mean stripe position, in the subset of embryos presenting all stripes (see methods).

**Figure 2: Spatiotemporal gap gene expression in gap gene mutants.**

(A–D) Simultaneously measured spatial gap gene expression profiles are shown for 85 embryos (42±4 min into nc14) in the *Kr* experiment for Kr (A), Kni (B), Hb (C), and Gt (D). Kr profiles are used for genotype assignment (see methods): homozygous (0X, magenta, n=29), hemizygous (1X, yellow, n=46), and quasi wild-type (2X, green, n=10). Dashed curves show the mean expression profile for each protein per genotype. (E) Top panel depicts the mean deviation of Gt in 0X-*Kr* ($\Delta I^{Gt}_{0X-Kr}$) from the wild-type pattern, defining the response to the removed *Kr* gene, which is estimated by subtracting the mean expression profile of Gt in 2X-*Kr* ($I^{Gt}_{2X-Kr}$) from that in 0X-*Kr* ($I^{Gt}_{0X-Kr}$). Bottom panel shows a color code for the same difference, where increased expression in the mutant relative to wild-type is indicated in red, and reduced expression is indicated in blue. (F–I) Spatiotemporal kymographs of color-coded rows with $\Delta I^{g}_{gt}$ as in E (bottom) and time running along the vertical axis from 10 to 56±4 min in nc14 for *Kr* (F), *kni* (G), *hb* (H), and *gt* (I) mutant genetic background experiments. The left column of each panel shows an example for a bootstrapped difference of resampled 2X data sets, estimating experimental noise in the $\Delta I$ measure (under ideal experimental conditions they would have been completely white). Middle and right column for each panel show $\Delta I$ measure for hemizygous and homozygous mutants, respectively. See S2 and methods for more information on measurement noise and expression variances.

**Figure S2: Measurement noise and expression variance of gap genes in gap mutants.**

(A) The mean spatiotemporal response of Gt in *0X-Kr*, *1X-Kr* and in three examples from resampled *2X-Kr* embryos. (B) Noise threshold for the minimal significant response, based on $\pm 2\sigma|_{Gt,2xKr}$, obtained by pooling n=200 resampled versions of $\Delta I^{Gt}_{2xKr}$. (C) Same as in A with grayed out pixels representing absolute response level lower than the $2\sigma$ threshold. (D) While the mean intensity of gap profiles in the gap mutants can show significant displacements, their coefficient of variation ($\sigma_i/g_i$, D top insets) is mostly kept near wild-type levels (green) for both homozygous (magenta) and hemizygous (yellow) gap mutants. Moreover, the relationship between the standard deviation $\sigma_i$ and the mean $g_i$ remains similar to wild-type (D, bottom insets). The positional error (E) measured for the 4 gap genes for the hemizygous genotypes (yellow), is also kept near wild-type levels (green). For the homozygous null genotypes (F), however the positional error ($\sigma_x 3g$: using the remaining 3 genes) show slight increase around the boundary region of the removed gene, compared with both wild-type measures for $\sigma_x 3g$ (blue) and $\sigma_x 4g$ (green). Nonetheless, at Eve stripes positions, the estimated positional error is bounded by ~2% of the egg length, compatible with that measured for eve stripes in the homozygous gap mutants. Note that for both homozygous and hemizygous mutans there is a single gene that provides similar positional accuracy to that obtained with all present gap genes (dashed black curve).

**Figure 3: Coexistence of code words with large dynamic range and reproducible patterning.**

(A) For each gap gene the mean expressions and associated standard deviations $\sigma_i$ (pooled over all expression profiles in the four 2X experiments, defining the code word expression range) is shown for each Eve expression peak (color code in inset) and trough (black) position. (B) Positional error along the AP axis: $\sigma_x^{4g}$ uses all 4 genes (gray circles), $\sigma_x^{1g}$ uses the single gap gene that minimizes positional error at each position (black dots). Note that at any position along the egg a single gap gene exists that can provide a comparable positional error to the one achieved with all four gap genes. (C) Main panel is as (B) with $\sigma_x^{1g}$ at Eve peaks (up-triangles) and troughs (down-triangles) with the color code indicating the gap gene providing the minimal $\sigma_x^{1g}$ at a given position. Top panel shows mean and standard deviation of gap gene expression profiles along the AP axis. Triangles annotate location and level of gap gene leading to minimal $\sigma_x^{1g}$. (D) Correlation between the local derivative dg/dx(x) and the code word range $\sigma_i(x)$ for the most accurate gap gene for inter peak/trough positions (black x); as well as for peaks (up-triangles), and troughs (down-triangles), with stripe identity indicated by eve color code (trough color matches its preceding peak). (E) 1st and 2nd principal components (PC) of all 2X code words at Eve peaks from individual embryos at 42±4 min, computed separately for each 2X experiment and overlayed together (color indicates stripe identity). For each dataset PC1 and PC2 include ~70% of the data variability. Eve peaks are perfectly separable across the pooled data, and troughs intervene between peaks with modest overlap. See also Figure S3.

**Figure S3: Sharp concentration changes compensate for wide dynamic range of code words, enabling reproducible patterning.**

(A) The cumulative probability of the standard deviations $\sigma_I$ of code word expression ranges at Eve peaks (green) and troughs (black) in the four individual 2X experiments (dashed lines), and mean across experiments (solid lines). 40% of peaks and troughs have $\sigma_i$ larger than 10% of the wild-type max. (B) The probability distribution of dg/dx(x) is shown for all gap genes (blue), and for the most accurate gap gene used in Figure 3C (orange). For the most accurate gene this distribution is shifted towards higher values. (C) The relationship between the coefficient of variation of expression level ($\sigma_i/<g_i>$) and its mean level ($<g_i>$) for the most accurate gene as measured in eve peaks, troughs and inter peak/trough positions (annotated by markers and eve color code as in Fig3D). (D) The relationship between the coefficient of variation of expression level ($\sigma_i/<g_i>$) and the local derivative ($dg/dx$) for the most accurate gene as measured in eve peaks, troughs and inter peak/trough positions. (E) PCA of Eve peak code words for each of the four 2X experiments. In individual experiments (genotype annotated) separability for peaks and troughs is enhanced. Note that for 2X-*gt*, stripes 1 and 2 are not completely separable (*gt* is on the X-chromosome and male embryos experience dosage compensation which blurs the code word separation capabilities).

**Figure 4: Hemizygous gap mutants only display local patterning effects.**

(A) Scatter plot of mean gap gene expression levels (at 42±4 min) for corresponding 1X versus 2X at Eve peak locations (as identified at 50±5 min, to enable network delay). For each 1X experiment there is one genetically manipulated gap gene with half wild-type dosage (red; 7*4 data points). Data not falling on the half-diagonal (y=0.5*x line) must have experienced some network compensatory effect; most data hover above the line shows compensation towards wild-type levels. For the unmanipulated gap genes, data points (black and green) consistently are on the unity line (within gray-shaded error), indicating wild-type levels. Some of these data correspond to Eve stripes that display significant positional shifts (green). For genotype breakdown see Figure S4A. (B) For green data in A, scatter plot of position shift Δx and expression level change for a single gap gene at the mutant stripe position. For each data point, marker edge color indicates genotype, marker shape indicates measured gene, and marker face color indicates Eve stripe identity. (C) Top shows kymograph of Kni expression in 1X-*Kr* experiment (from Figure 2F), middle shows Eve pattern for 1X-*Kr* (from Figure S1B), and the bottom inset shows the displacement and positional error of each stripe (from Figure S1B). Stripe 5 shows a significantly displaced mean position, highlighted by yellow dashed line. Line falls on red region in kymograph, indicating increased Kni expression to maintain a wild-type *eve*-activating level. A milder similar effect is observed for stripe 4, highlighted by green dashed line; as well as for stripe 6, for which Gt is adjusting its level (see Panel B). (F) Projection of gap gene code words for Eve peaks from all hemizygous mutant embryos (squares, color-coded by stripe) onto the PC1–PC2 plane defined by wild-type Eve peaks (black triangles) at 42±4 min. Code words encoding different Eve peaks are clearly separable, yet their extend is wider due to near 50% reduction in the manipulated gene. See Figure S4 for separability of hemizygous code words for peaks and troughs.

**Figure S4: Hemizygous gap mutants actively maintain near wild-type patterning.**

(A-D) As Figure 4A for individual 1X genotypes, as annotated in upper left of each panel. All hemizygous mutants are showing conserved concentrations for all unperturbed gap proteins. (E) We estimate the response per position, $R_{gtp}^g(x)$, by taking the square-root of the averaged squared mean deviation ($\Delta I_{gtp}^g(x,t)$) across time (see methods). For each gap gene (see color code) $R_{gtp}^g(x)$ is shown for the hemizygous mutants (1X experiments, solid lines) and for the bootstrapped 2X experiments (dashed lines) along the AP axis. Note that the overall spatial extend of $R_{gtp}^g(x)$ is larger for the unperturbed than that for the perturbed gene domains. (F) $\chi^2$-per gene is shown for code words from individual embryos (gray) along the AP axis (dashed black curve designates mean across embryos). All code words in hemizygous embryos show distance lower than the maximal distance (horizontal dashed line) of any wild-type code word from its distribution across wild-type embryos. While conserved code words are observed throughout the embryo, the unperturbed genes are changing their levels within and outside the perturbed gene domain. Together this data (E-F) generalize the example shown in Figure 4C-E by suggesting that conserved code words along the AP axis often stem from active work of the gap gene network, and particularly its unperturbed genes, rather than trivial conservation due to no change in expression levels. (G) While some code words are altered (by nearly halving the levels of the perturbed gene), such that gap code words in hemizygous mutants occupy a wider region than their wild-type version in the PC axes, peaks (squares) and troughs (triangles) of hemizygous code words are still separable.

**Figure 5: Canonical and non-canonical code words in homozygous mutants.**

(A) Top panel shows expression patterns of Eve in individual embryos (solid gray) and the mean Eve pattern (magenta) in 0X-*Kr* experiment; graph is overlayed with the mean 2X-*Kr* Eve pattern (green) and the mean wild-type expression pattern of Kr (dashed gray). Circles indicate peak position and Eve intensity in individual embryos. Roman numbers indicate the serial position of a peak along the AP axis. Percentages indicate partial stripe penetrance if lower than 97%. Bottom panel shows the projections of gap code words at Eve peaks (black) for 0X-*Kr* experiment onto the PC1–PC2 plane obtained for gap code-words at Eve peaks in 2x-*Kr* (wild-type). Percentages on PC axes indicate the proportion of the wild-type variance explained by a particular PC. (B–D) Similar to A but for 0X-*kni* (B), 0X-*gt* (C) and 0X-*hb* (D) experiments, respectively. Mutant stripe identities can be inferred by their overlap with the wild-type stripe cloud (see Eve color code on left side) and/or proximity to more than one stripe. For instance, code words activating *eve* in mutant stripe II in A occupy the region proximal to Eve stripe 2 but it is also close to stripe 1 and 7. Eve reporter experiments qualitatively suggest that indeed three separate *eve* enhancers are active around stripe II. Similar conclusions can be drawn for 0X-*kni* (B), 0X-*gt* (C) and 0X-*hb* (D) experiments (see text for stripe identity interpretation).

**Figure S5: Gap network induced active concentration adjustments in homozygous mutants.**

(A) $\chi^2$-per gene (gray) estimates the distance between code words from individual homozygous null embryos (0X-*hb*, 0X-*Kr*, 0X-*gt,* and 0X-*kni* from left to right) at a given position and the reference distribution of code words spanned by wild-type embryos (see methods). Black dashed curves show the mean $\chi^2$-per gene. The wild-type expression profile of the removed gap gene is overlayed, as well as the mean Eve stripe positions for the homozygous null mutants (vertical magenta lines). Note that $\chi^2$ levels are only elevated above baseline in regions of the embryo that overlap with the removed gene domain. Only ~10% of 0X code words across the 0.1–0.9 EL span are more distal than the maximal distance measured for a wild-type code word from its distribution across embryos (horizontal dashed lines). (B) $R_{gt}^g(x)$ is shown per gene (gap gene color code) and genotype (same order as in A) indicating a significant response (solid lines, color code denotes gap genes) along the main trunk, compared with $R_{gt}^g(x)$ obtained from the bootstrapped 2X experiments (dashed lines). Note that the spatial range where potentially novel code words appear in A is much more restricted, indicating that the response in most embryo regions is *actively* drawn to the familiar wild-type repertoire of code words. (C) An example for such active adjustments in 0X-*Kr* embryos where Eve stripes II–V (top panel) are displaced yet gap gene levels are actively adapted at the new Eve stripe positions (bottom panel), such that the levels observed in each of the 5 mutant stripes can be traced back to a wild-type stripe with strong compatibility with wild-type levels for all gap genes.

**Figure 1.**

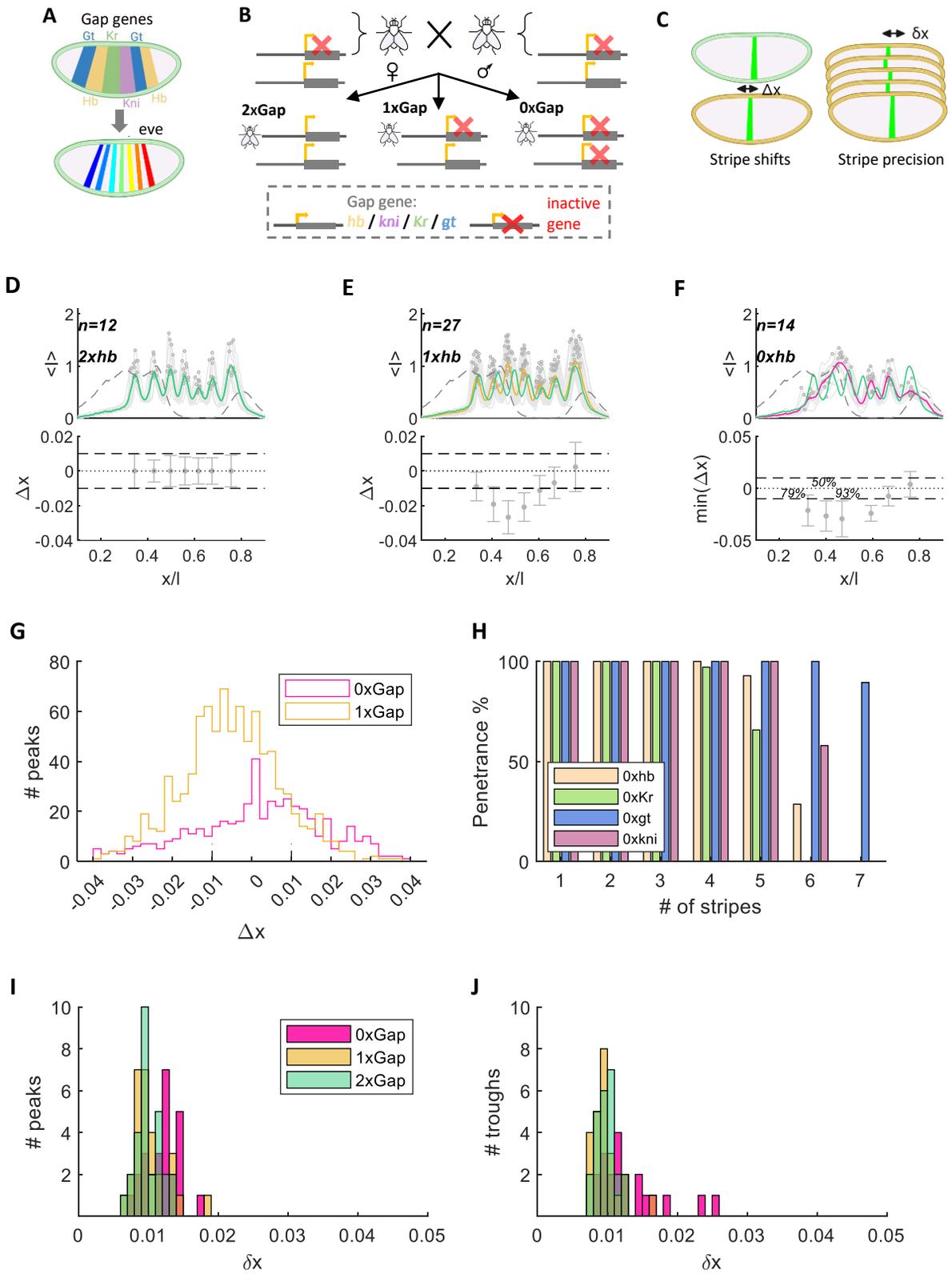

**Figure S1.**

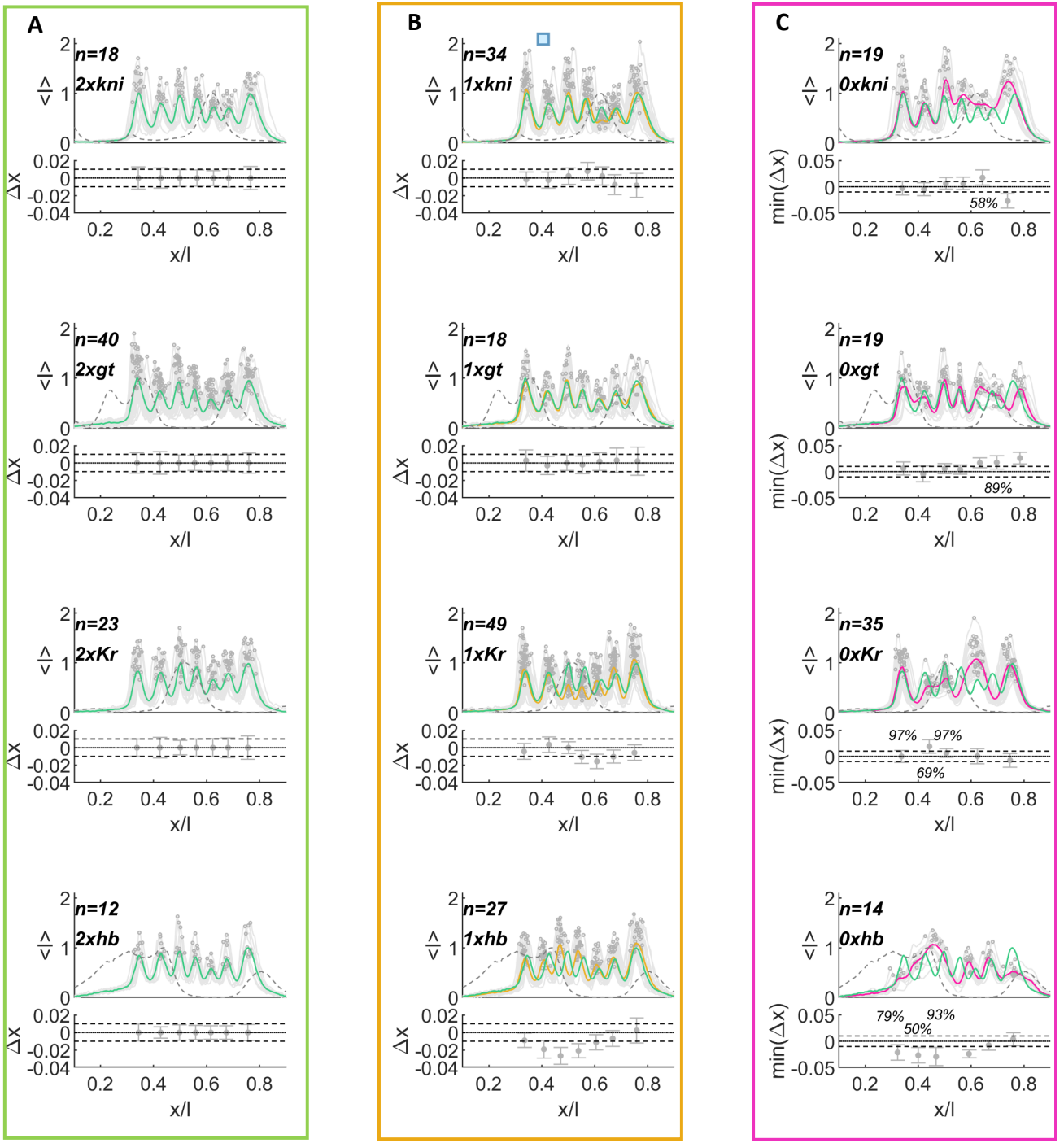
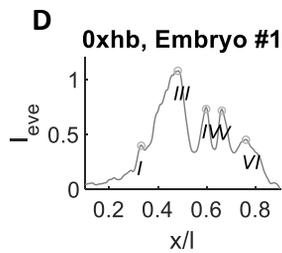
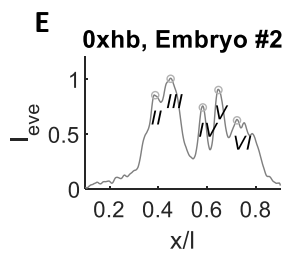
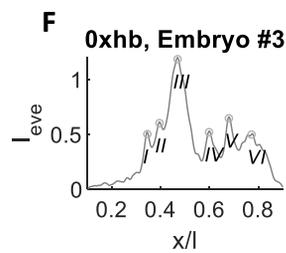

**Figure 2.**

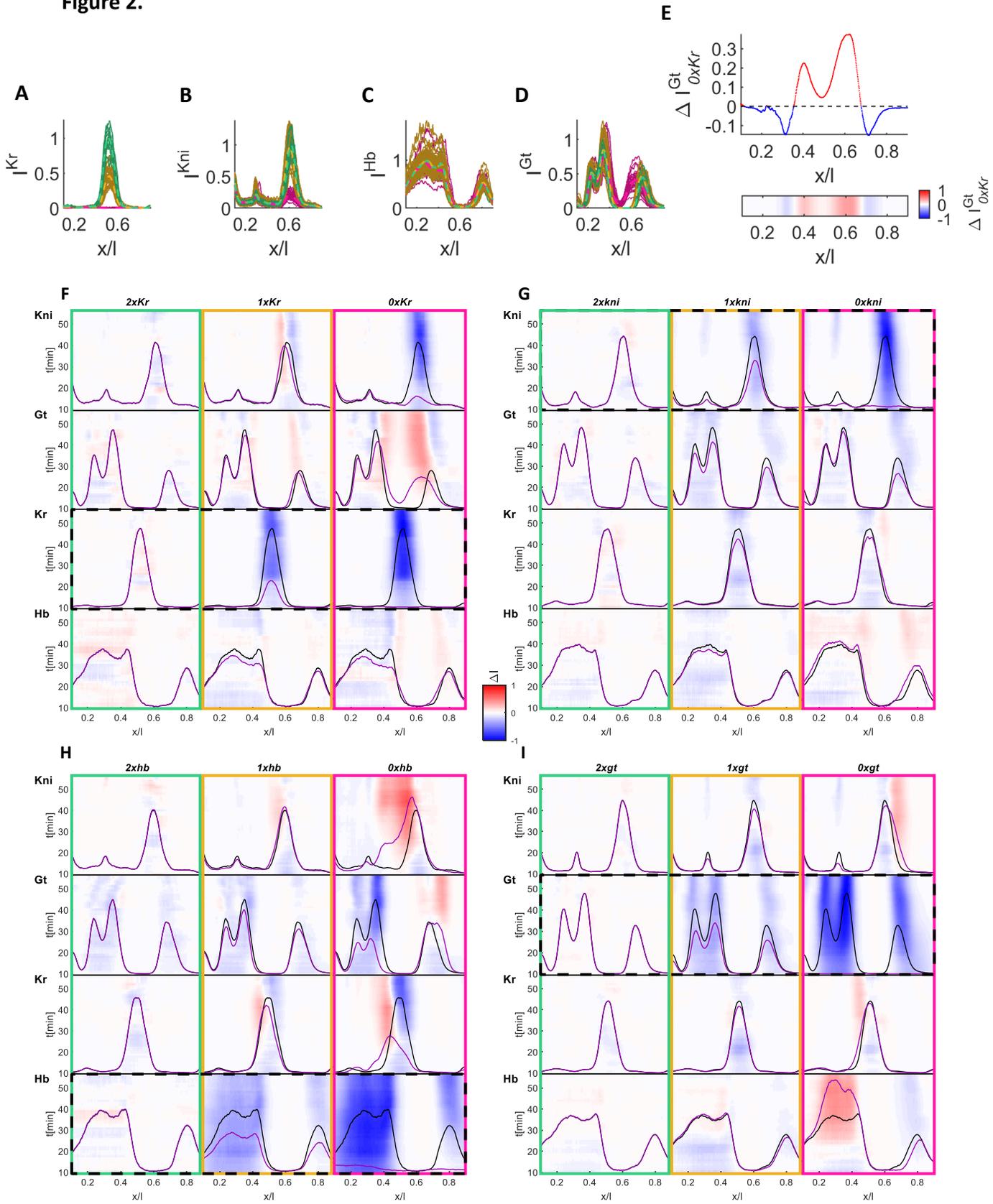

**Figure S2.**

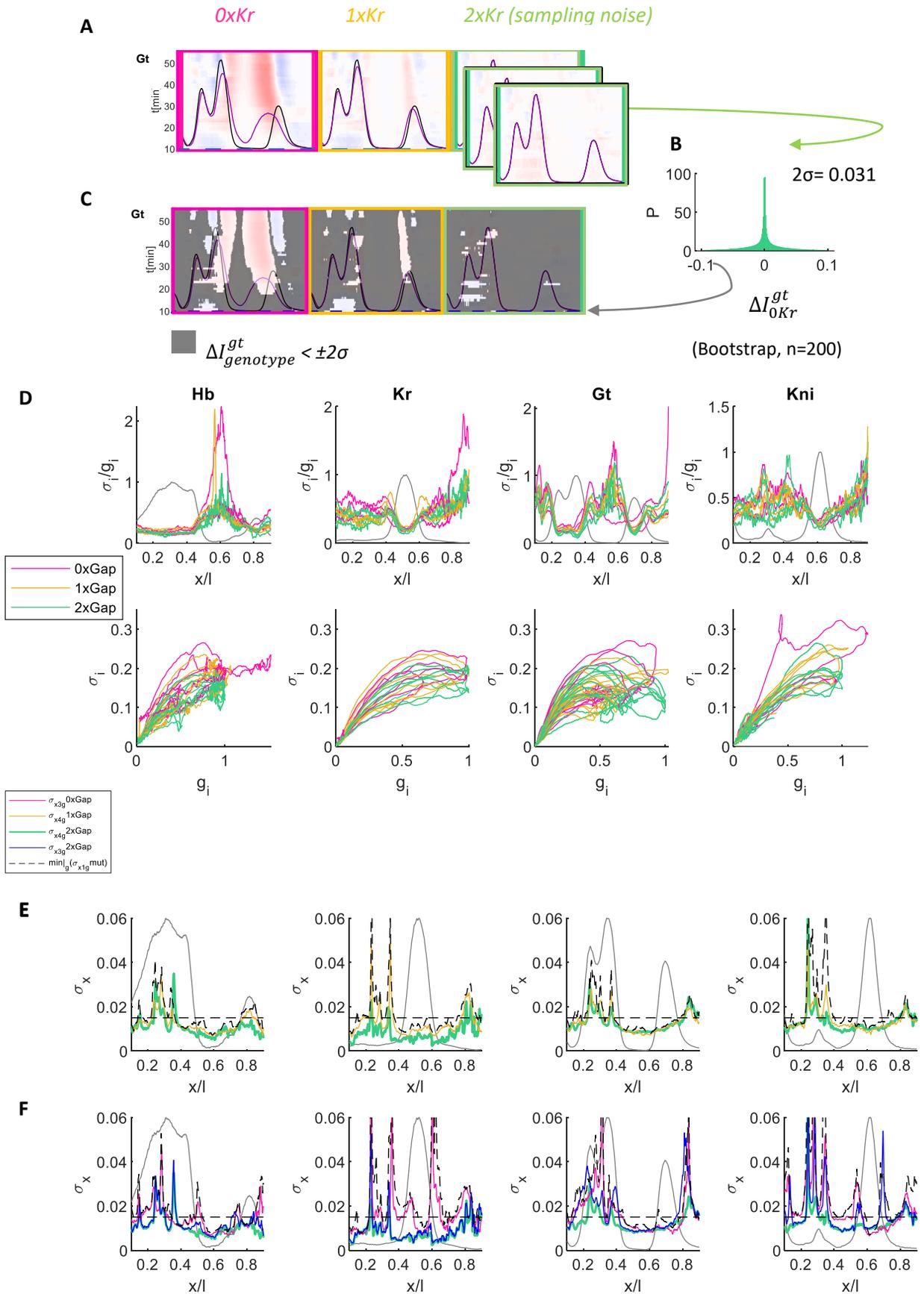

**Figure 3.**

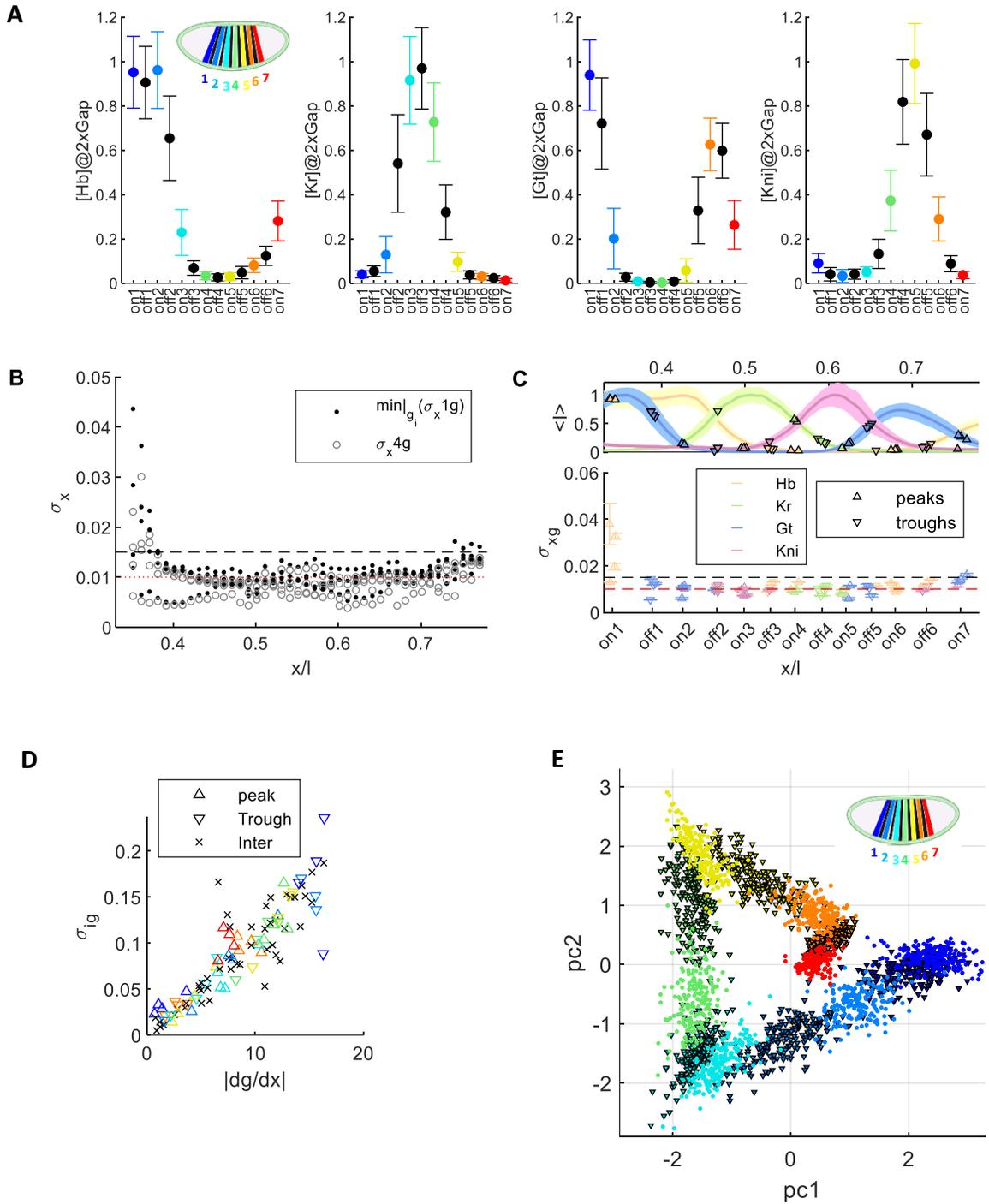

**Figure S3.**

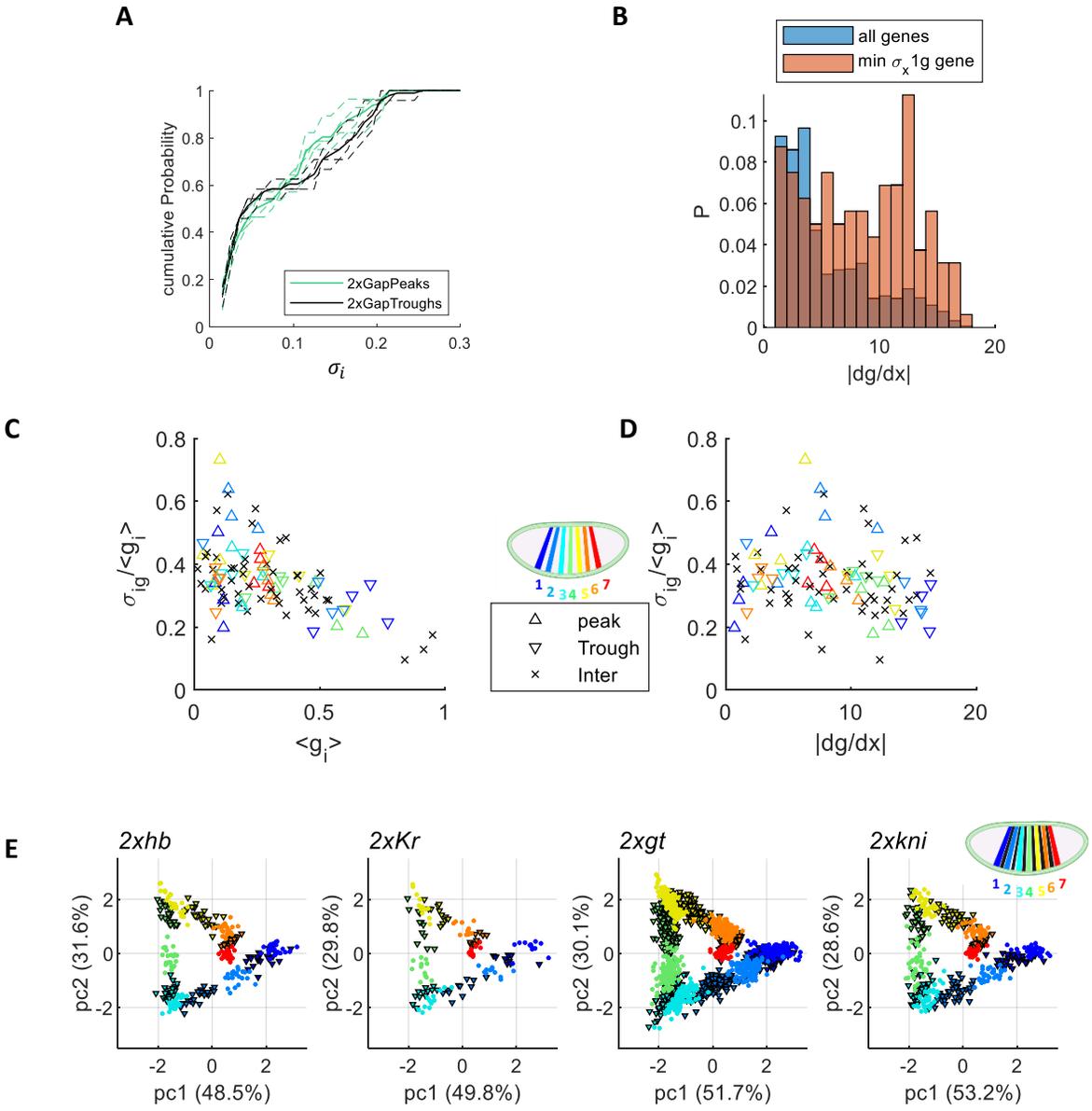

**Figure 4.**

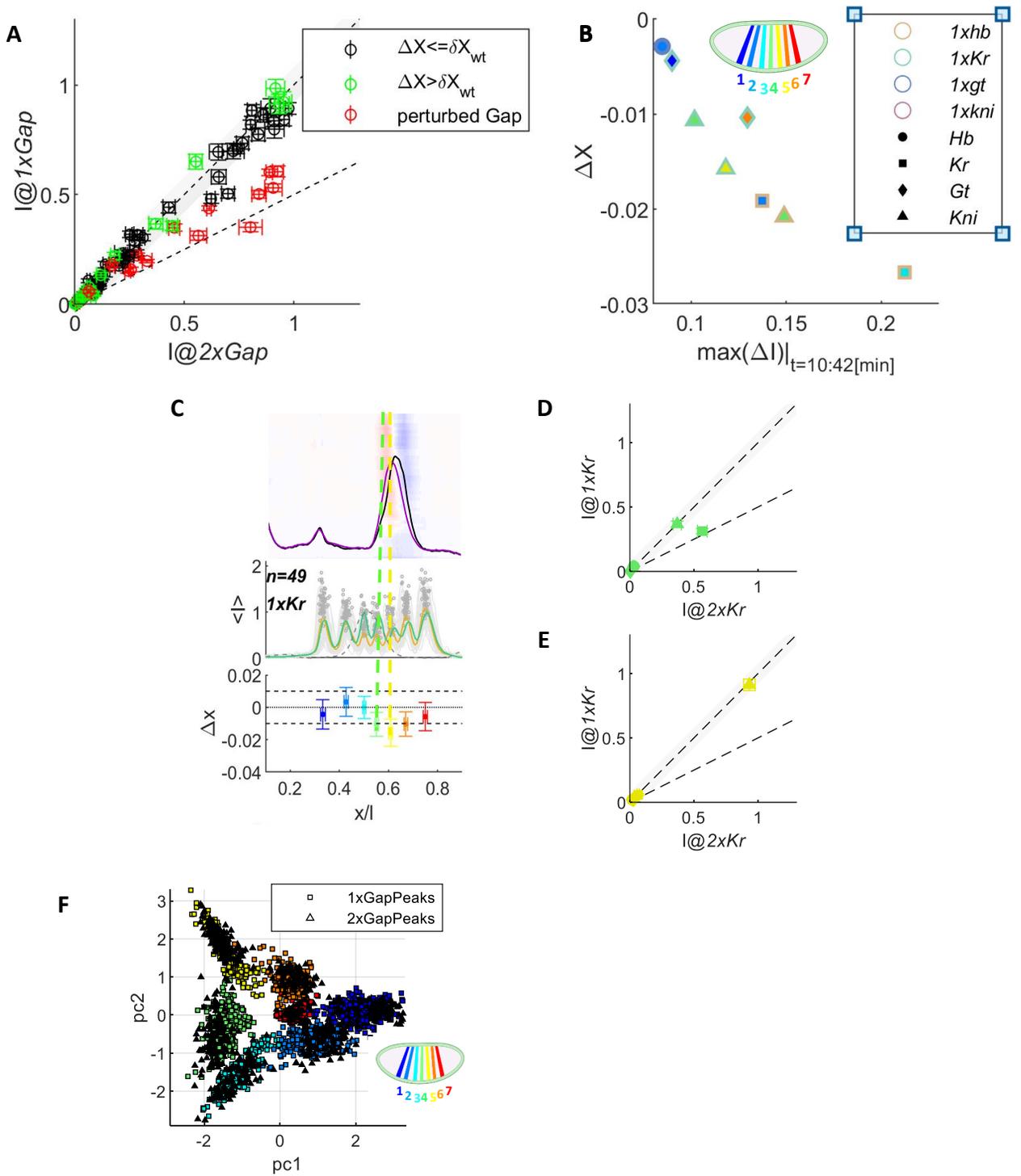

**Figure S4.**

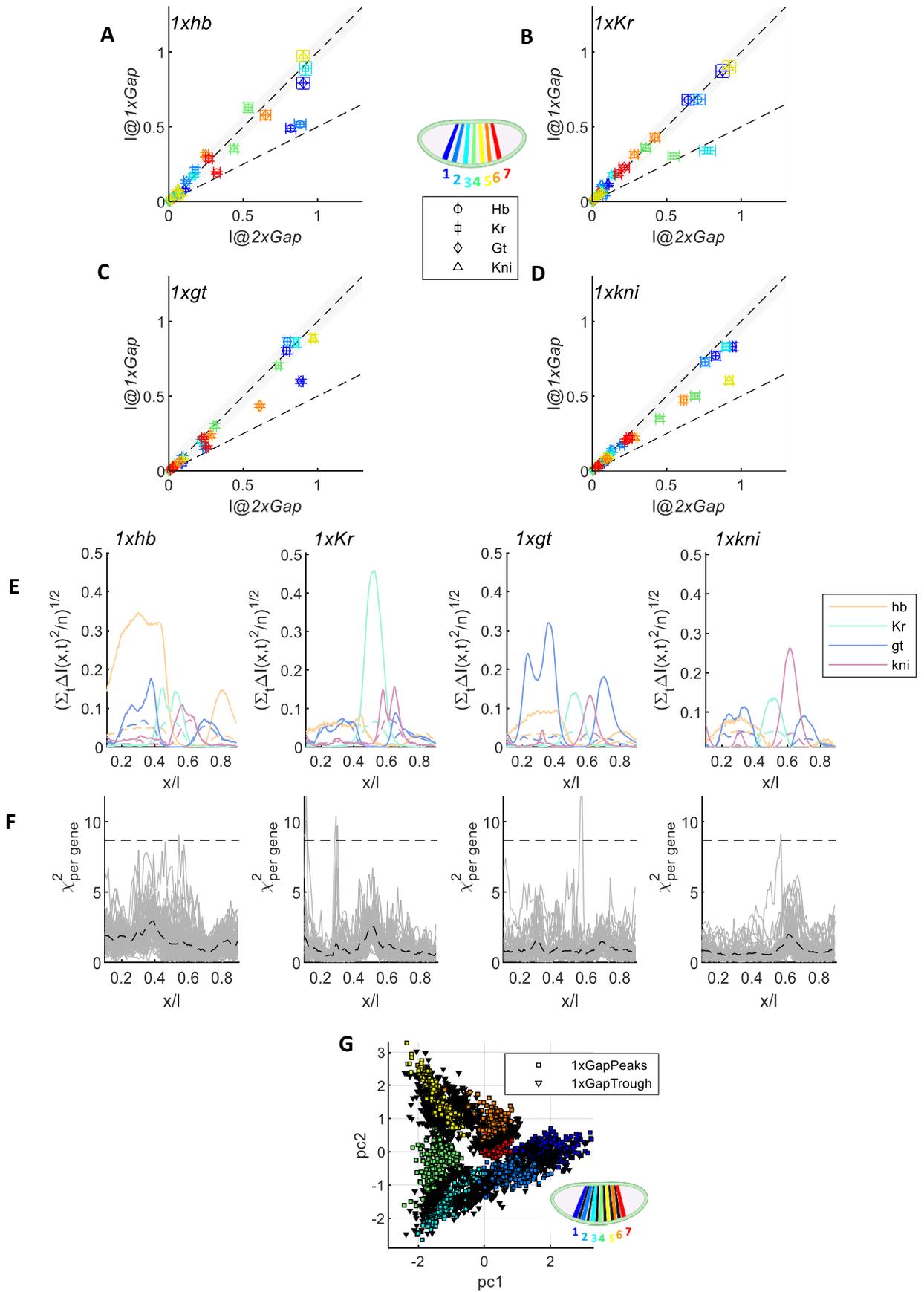

**Figure 5.**

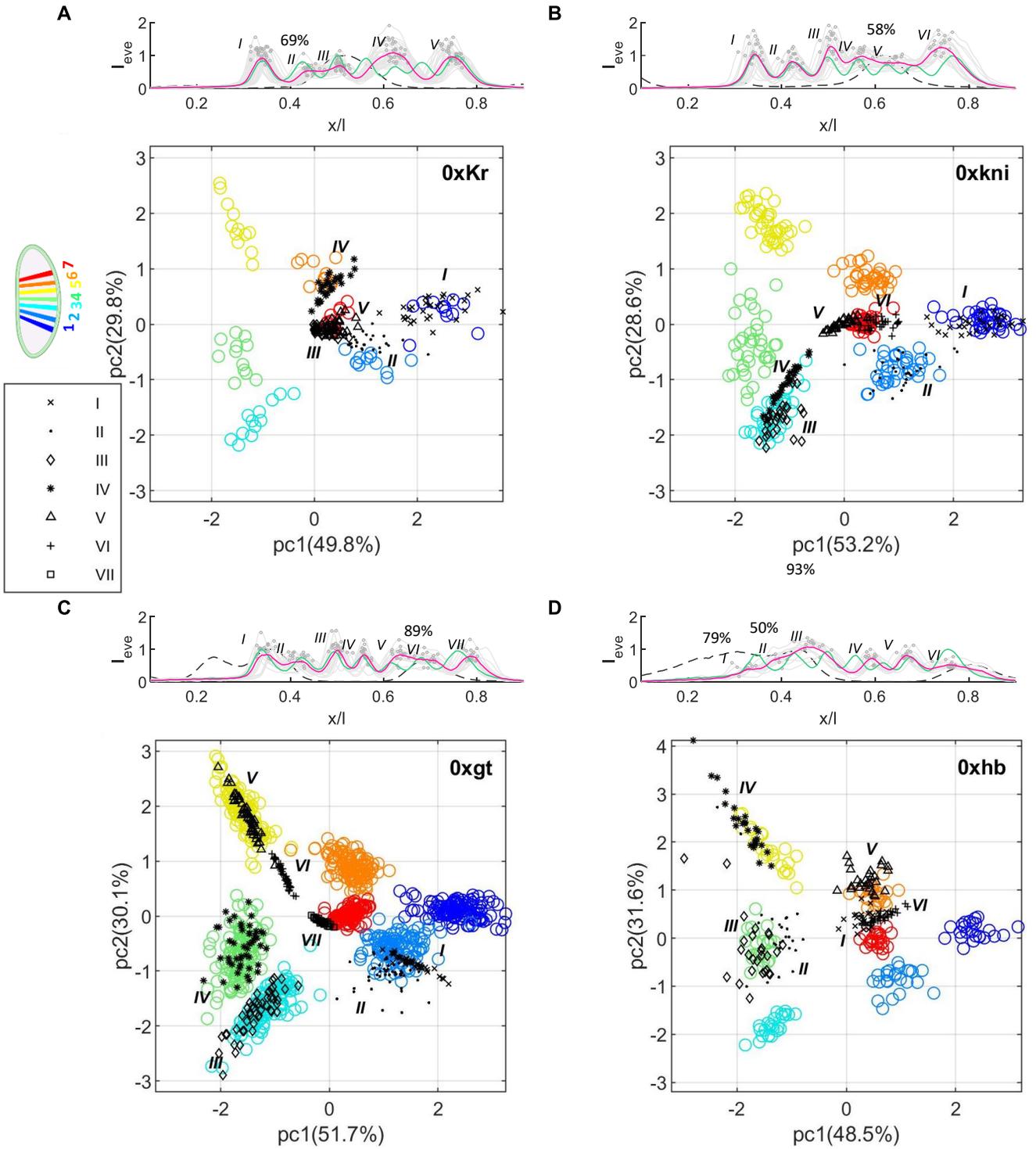

**Figure S5.**

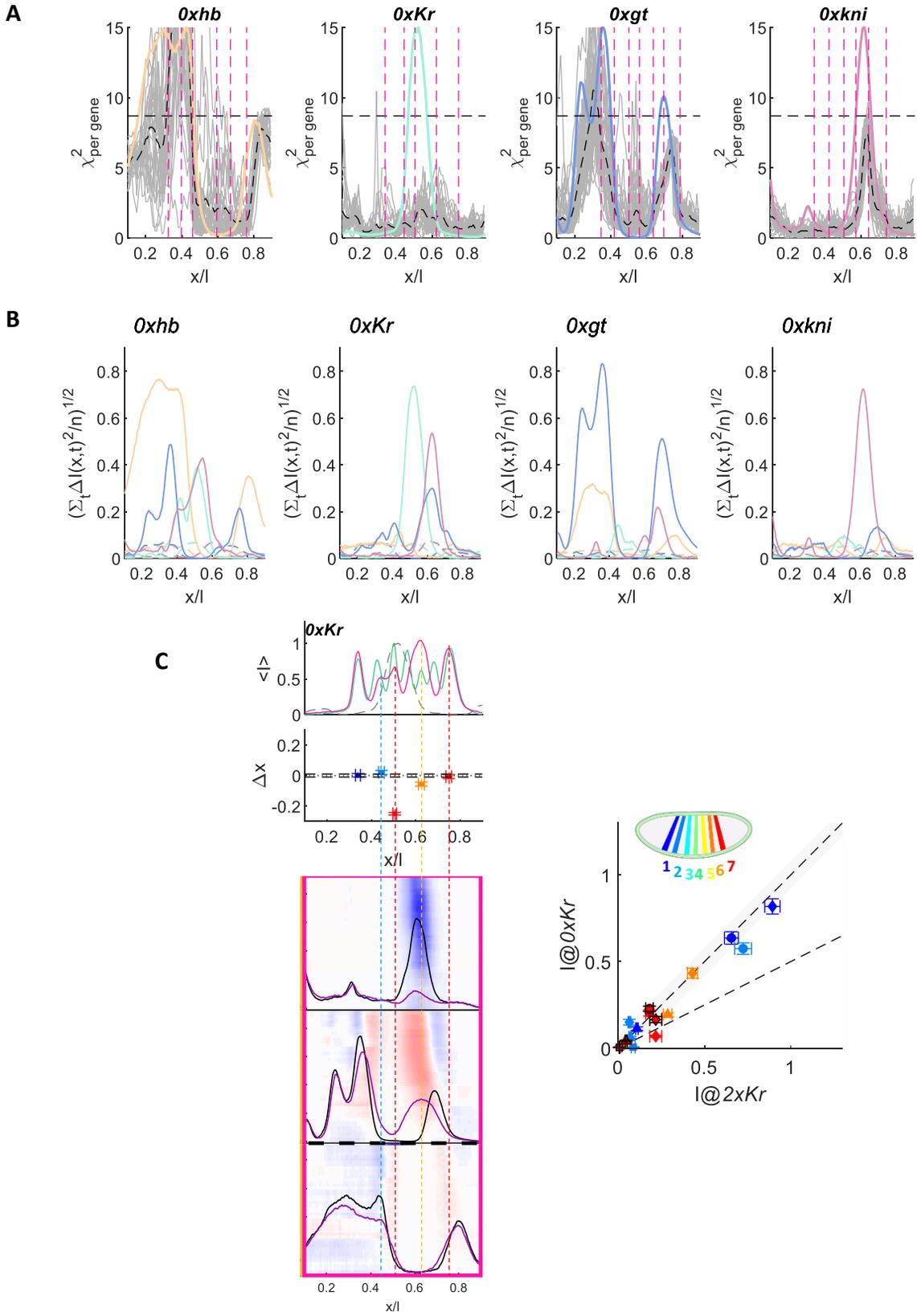